\documentclass[prl,superscriptaddress,twocolumn,showpacs,amsmath,amssymb,preprintnumbers]{revtex4-1}

\usepackage{hyperref}
\usepackage{dcolumn}
\usepackage{bm}
\usepackage{ulem}
\usepackage{xspace}
\usepackage{graphicx}

\newcommand {\pbi} {{\rm pb}^{-1}}
\newcommand {\ppbar}{p\bar{p}}
\newcommand {\Et}   {\ensuremath{E_{\mathrm{T}}}}
\newcommand {\vecEt}{\ensuremath{\vec{E}_{\mathrm{T}}}}
\newcommand {\mll}  {m_{\ell\ell}}

\chardef\letterchar=11
\chardef\otherchar=12
\chardef\eolinechar=5

\newcount\hrs\newcount\minu\newcount\temptime
\def\hm{\hrs=\time \divide\hrs by 60 \minu=\time\temptime=\hrs
\multiply\temptime by 60
\advance\minu by -\temptime
\ifnum\minu<10 \let\zerofill=0\else \let\zerofill=\relax\fi
 \the\hrs:\zerofill\the\minu}

\def\lapprox{\ensuremath{\sim\kern-1em\raise 0.65ex\hbox{$<$}}}
\def\rapprox{\ensuremath{\sim\kern-1em\raise 0.65ex\hbox{$>$}}}

\def\antibar#1{\ensuremath{#1\bar{#1}}}

\def\ttbar{\antibar{t}}

\def\qqbar{\antibar{q}}

\def\Wplus{\ensuremath{W^+}}
\def\Wminus{\ensuremath{W^-}}

\def\chic{\ensuremath{\raise.4ex\hbox{$\chi$}_{{c}}}}

\def\chib{\ensuremath{\raise.4ex\hbox{$\chi$}_{{b}}}}

\let\psii=\psi  
\def\psi{\ensuremath{\psii}}

\def\ggino{\ensuremath{\mathchoice
      {\displaystyle\raise.4ex\hbox{$\displaystyle\tilde\chi$}}
         {\textstyle\raise.4ex\hbox{$\textstyle\tilde\chi$}}
       {\scriptstyle\raise.3ex\hbox{$\scriptstyle\tilde\chi$}}
 {\scriptscriptstyle\raise.3ex\hbox{$\scriptscriptstyle\tilde\chi$}}}}

\def\chinop{\ensuremath{\mathchoice
      {\displaystyle\raise.4ex\hbox{$\displaystyle\tilde\chi^+$}}
         {\textstyle\raise.4ex\hbox{$\textstyle\tilde\chi^+$}}
       {\scriptstyle\raise.3ex\hbox{$\scriptstyle\tilde\chi^+$}}
 {\scriptscriptstyle\raise.3ex\hbox{$\scriptscriptstyle\tilde\chi^+$}}}}
\def\chinom{\ensuremath{\mathchoice
      {\displaystyle\raise.4ex\hbox{$\displaystyle\tilde\chi^-$}}
         {\textstyle\raise.4ex\hbox{$\textstyle\tilde\chi^-$}}
       {\scriptstyle\raise.3ex\hbox{$\scriptstyle\tilde\chi^-$}}
 {\scriptscriptstyle\raise.3ex\hbox{$\scriptscriptstyle\tilde\chi^-$}}}}
\def\chinopm{\ensuremath{\mathchoice
      {\displaystyle\raise.4ex\hbox{$\displaystyle\tilde\chi^\pm$}}
         {\textstyle\raise.4ex\hbox{$\textstyle\tilde\chi^\pm$}}
       {\scriptstyle\raise.3ex\hbox{$\scriptstyle\tilde\chi^\pm$}}
 {\scriptscriptstyle\raise.3ex\hbox{$\scriptscriptstyle\tilde\chi^\pm$}}}}
\def\chinomp{\ensuremath{\mathchoice
      {\displaystyle\raise.4ex\hbox{$\displaystyle\tilde\chi^\mp$}}
         {\textstyle\raise.4ex\hbox{$\textstyle\tilde\chi^\mp$}}
       {\scriptstyle\raise.3ex\hbox{$\scriptstyle\tilde\chi^\mp$}}
 {\scriptscriptstyle\raise.3ex\hbox{$\scriptscriptstyle\tilde\chi^\mp$}}}}

\def\chinoonep{\ensuremath{\mathchoice
      {\displaystyle\raise.4ex\hbox{$\displaystyle\tilde\chi^+_1$}}
         {\textstyle\raise.4ex\hbox{$\textstyle\tilde\chi^+_1$}}
       {\scriptstyle\raise.3ex\hbox{$\scriptstyle\tilde\chi^+_1$}}
 {\scriptscriptstyle\raise.3ex\hbox{$\scriptscriptstyle\tilde\chi^+_1$}}}}
\def\chinoonem{\ensuremath{\mathchoice
      {\displaystyle\raise.4ex\hbox{$\displaystyle\tilde\chi^-_1$}}
         {\textstyle\raise.4ex\hbox{$\textstyle\tilde\chi^-_1$}}
       {\scriptstyle\raise.3ex\hbox{$\scriptstyle\tilde\chi^-_1$}}
 {\scriptscriptstyle\raise.3ex\hbox{$\scriptscriptstyle\tilde\chi^-_1$}}}}
\def\chinoonepm{\ensuremath{\mathchoice
      {\displaystyle\raise.4ex\hbox{$\displaystyle\tilde\chi^\pm_1$}}
         {\textstyle\raise.4ex\hbox{$\textstyle\tilde\chi^\pm_1$}}
       {\scriptstyle\raise.3ex\hbox{$\scriptstyle\tilde\chi^\pm_1$}}
 {\scriptscriptstyle\raise.3ex\hbox{$\scriptscriptstyle\tilde\chi^\pm_1$}}}}

\def\chinotwop{\ensuremath{\mathchoice
      {\displaystyle\raise.4ex\hbox{$\displaystyle\tilde\chi^+_2$}}
         {\textstyle\raise.4ex\hbox{$\textstyle\tilde\chi^+_2$}}
       {\scriptstyle\raise.3ex\hbox{$\scriptstyle\tilde\chi^+_2$}}
 {\scriptscriptstyle\raise.3ex\hbox{$\scriptscriptstyle\tilde\chi^+_2$}}}}
\def\chinotwom{\ensuremath{\mathchoice
      {\displaystyle\raise.4ex\hbox{$\displaystyle\tilde\chi^-_2$}}
         {\textstyle\raise.4ex\hbox{$\textstyle\tilde\chi^-_2$}}
       {\scriptstyle\raise.3ex\hbox{$\scriptstyle\tilde\chi^-_2$}}
 {\scriptscriptstyle\raise.3ex\hbox{$\scriptscriptstyle\tilde\chi^-_2$}}}}
\def\chinotwopm{\ensuremath{\mathchoice
      {\displaystyle\raise.4ex\hbox{$\displaystyle\tilde\chi^\pm_2$}}
         {\textstyle\raise.4ex\hbox{$\textstyle\tilde\chi^\pm_2$}}
       {\scriptstyle\raise.3ex\hbox{$\scriptstyle\tilde\chi^\pm_2$}}
 {\scriptscriptstyle\raise.3ex\hbox{$\scriptscriptstyle\tilde\chi^\pm_2$}}}}

\def\nino{\ensuremath{\mathchoice
      {\displaystyle\raise.4ex\hbox{$\displaystyle\tilde\chi^0$}}
         {\textstyle\raise.4ex\hbox{$\textstyle\tilde\chi^0$}}
       {\scriptstyle\raise.3ex\hbox{$\scriptstyle\tilde\chi^0$}}
 {\scriptscriptstyle\raise.3ex\hbox{$\scriptscriptstyle\tilde\chi^0$}}}}

\def\ninoone{\ensuremath{\mathchoice
      {\displaystyle\raise.4ex\hbox{$\displaystyle\tilde\chi^0_1$}}
         {\textstyle\raise.4ex\hbox{$\textstyle\tilde\chi^0_1$}}
       {\scriptstyle\raise.3ex\hbox{$\scriptstyle\tilde\chi^0_1$}}
 {\scriptscriptstyle\raise.3ex\hbox{$\scriptscriptstyle\tilde\chi^0_1$}}}}
\def\ninotwo{\ensuremath{\mathchoice
      {\displaystyle\raise.4ex\hbox{$\displaystyle\tilde\chi^0_2$}}
         {\textstyle\raise.4ex\hbox{$\textstyle\tilde\chi^0_2$}}
       {\scriptstyle\raise.3ex\hbox{$\scriptstyle\tilde\chi^0_2$}}
 {\scriptscriptstyle\raise.3ex\hbox{$\scriptscriptstyle\tilde\chi^0_2$}}}}
\def\ninothree{\ensuremath{\mathchoice
      {\displaystyle\raise.4ex\hbox{$\displaystyle\tilde\chi^0_3$}}
         {\textstyle\raise.4ex\hbox{$\textstyle\tilde\chi^0_3$}}
       {\scriptstyle\raise.3ex\hbox{$\scriptstyle\tilde\chi^0_3$}}
 {\scriptscriptstyle\raise.3ex\hbox{$\scriptscriptstyle\tilde\chi^0_3$}}}}
\def\ninofour{\ensuremath{\mathchoice
      {\displaystyle\raise.4ex\hbox{$\displaystyle\tilde\chi^0_4$}}
         {\textstyle\raise.4ex\hbox{$\textstyle\tilde\chi^0_4$}}
       {\scriptstyle\raise.3ex\hbox{$\scriptstyle\tilde\chi^0_4$}}
 {\scriptscriptstyle\raise.3ex\hbox{$\scriptscriptstyle\tilde\chi^0_4$}}}}

\let\pii=\pi
\def\pi{\ensuremath{\pii}}

\let\etaa=\eta
\def\eta{\ensuremath{\etaa}}

\def\pt{\ensuremath{p_{\mathrm{T}}}} 
\def\pT{\ensuremath{p_{\mathrm{T}}}} 
\def\et{\ensuremath{E_{\mathrm{T}}}} 
 
\def\ET{\ensuremath{E_{\mathrm{T}}}}

\def\met{\ensuremath{E_{\mathrm{T}}^{\mathrm{miss}}}}

\def\TeV{\ifmmode {\mathrm{\ Te\kern -0.1em V}}\else
                   \textrm{Te\kern -0.1em V}\fi}
\def\GeV{\ifmmode {\mathrm{\ Ge\kern -0.1em V}}\else
                   \textrm{Ge\kern -0.1em V}\fi}
\def\MeV{\ifmmode {\mathrm{\ Me\kern -0.1em V}}\else
                   \textrm{Me\kern -0.1em V}\fi}
\def\keV{\ifmmode {\mathrm{\ ke\kern -0.1em V}}\else
                   \textrm{ke\kern -0.1em V}\fi}
\def\eV{\ifmmode  {\mathrm{\ e\kern -0.1em V}}\else
                   \textrm{e\kern -0.1em V}\fi}

\def\TeVc{\ifmmode {\mathrm{\ Te\kern -0.1em V}/c}\else
                   {\textrm{Te\kern -0.1em V}/$c$}\fi}
\def\GeVc{\ifmmode {\mathrm{\ Ge\kern -0.1em V}/c}\else
                   {\textrm{Ge\kern -0.1em V}/$c$}\fi}
\def\MeVc{\ifmmode {\mathrm{\ Me\kern -0.1em V}/c}\else
                   {\textrm{Me\kern -0.1em V}/$c$}\fi}
\def\keVc{\ifmmode {\mathrm{\ ke\kern -0.1em V}/c}\else
                   {\textrm{ke\kern -0.1em V}/$c$}\fi}
\def\eVc{\ifmmode  {\mathrm{\ e\kern -0.1em V}/c}\else
                   {\textrm{e\kern -0.1em V}/$c$}\fi}

\def\TeVcc{\ifmmode {\mathrm{\ Te\kern -0.1em V}/c^2}\else
                   {\textrm{Te\kern -0.1em V}/$c^2$}\fi}
\def\GeVcc{\ifmmode {\mathrm{\ Ge\kern -0.1em V}/c^2}\else
                   {\textrm{Ge\kern -0.1em V}/$c^2$}\fi}
\def\MeVcc{\ifmmode {\mathrm{\ Me\kern -0.1em V}/c^2}\else
                   {\textrm{Me\kern -0.1em V}/$c^2$}\fi}
\def\keVcc{\ifmmode {\mathrm{\ ke\kern -0.1em V}/c^2}\else
                   {\textrm{ke\kern -0.1em V}/$c^2$}\fi}
\def\eVcc{\ifmmode  {\mathrm{\ e\kern -0.1em V}/c^2}\else
                   {\textrm{e\kern -0.1em V}/$c^2$}\fi}

\def\cm{\ifmmode  {\mathrm{\ cm}}\else
                   \textrm{~cm}\fi}

\newbox\boxsqbox
\newdimen\boxsize\boxsize=1.2ex
\def\boxop{
\setbox\boxsqbox=\vbox{\hrule depth0.8pt width0.8\boxsize height0pt
                       \kern0.8\boxsize
                       \hrule height0.8pt width0.8\boxsize depth0pt}
           \hbox{
           \vrule height1.0\boxsize width0.8pt depth0pt
           \copy\boxsqbox
           \vrule height1.0\boxsize width0.8pt depth0pt\kern1.5pt}}

\def\pmb#1{\setbox0=\hbox{$#1$}
  \kern-.025em\copy0\kern-1.0\wd0
  \kern.05em\copy0\kern-1.0\wd0
  \kern-.025em\raise.0433em\box0}

\newdimen\dkwidth
\def\dk{
   \dkwidth=\baselineskip
   {\def\to{\rightarrow}
   \kern 3pt
   \hbox{
      \raise 3pt
      \hbox{
         \vrule height 0.8\dkwidth width 0.7pt depth0pt
      }
      \kern-0.4pt
      \hbox to 1.5\dkwidth{
         \rightarrowfill
      }
   \kern0.6em
   }}
}

\def\eqalign#1{
   \,
   \vcenter{
      \openup\jot\m@th
      \ialign{
         \strut\hfil$\displaystyle{##}$&&$
         \displaystyle{{}##}$\hfil\crcr#1\crcr
      }
   }
   \,
}

\def\mythicksim{\ensuremath{\thicksim\kern -0.3em}}

\newcommand{\sla}[1]{{\raise.15ex\hbox{$/$}\kern-.67em #1}}

\def\ord{\ifmmode{\mathcal{O}}\else$\mathcal{O}$\fi}
\def\cov{\ifmmode{\mathcal{C}}\else$\mathcal{C}$\fi}
\def\lum{\ifmmode{\mathcal{L}}\else$\mathcal{L}$\fi}
\def\lik{\ifmmode{\mathcal{L}}\else$\mathcal{L}$\fi}
\def\likr{\ifmmode{\mathcal{LR}}\else$\mathcal{LR}$\fi}
\def\br{\ifmmode{\mathcal{BR}}\else$\mathcal{BR}$\fi}
\def\dil{\ifmmode{\mathcal{D}}\else$\mathcal{D}$\fi}
\def\amp{\ifmmode{\mathcal{A}}\else$\mathcal{A}$\fi}
\def\asy{\ifmmode{\mathcal{A}}\else$\mathcal{A}$\fi}
\def\sig{\ifmmode{\mathcal{S}}\else$\mathcal{S}$\fi}
\def\bkg{\ifmmode{\mathcal{B}}\else$\mathcal{B}$\fi}
\def\gaus{\ifmmode{\mathcal{G}}\else$\mathcal{G}$\fi}
\def\kappa{\ensuremath{k}}

\def\d0sig{\ensuremath{d_0/\sigma_{d_0}}}

\def\pt{\ensuremath{p_{\mathrm{T}}}\xspace}

\def\D{\ensuremath{\mathrm{D}}}

\def\Z0{\ensuremath{\mathrm{Z}^0}}

\def\Phi{\ensuremath{\phi^0}}

\def\RhoPiPi0{\ensuremath{\rho^+ \to \pi^+\pi^0}}

\def\WW{\ensuremath{\Wplus\Wminus}\xspace}

\def\pythia{{\sc pythia}\xspace}

\def\mcatnlo{{\sc mc@nlo}\xspace}
\def\alpgen{{\sc alpgen}\xspace}
\def\herwig{{\sc herwig}\xspace}
\def\Jimmy{{\sc jimmy}\xspace}
\def\geant{{\sc geant4}\xspace}

\def\ggtwoww{{\sc gg2ww}\xspace}
\def\ptsp1{\ensuremath{p_{\mathrm{T}}} }

\def\met{\ensuremath{E_{\mathrm{T}}^{\mathrm{miss}}}\xspace}
\def\metvec{\ensuremath{\vec{E}_{\mathrm{T}}^{\mathrm{miss}}}\xspace}

\def\metrel{\ensuremath{E_{\mathrm{T,rel}}^{\mathrm{\mathrm{miss}}}}\xspace}

\def\TeV{\ensuremath{\mathrm{Te\kern -0.1em V}}\xspace}
\def\GeV{\ensuremath{\mathrm{Ge\kern -0.1em V}}\xspace}
\def\MeV{\ensuremath{\mathrm{Me\kern -0.1em V}}\xspace}
\def\keV{\ensuremath{\mathrm{ke\kern -0.1em V}}\xspace}

\def\TeVc{\ensuremath{\mathrm{Te\kern -0.1em V}\kern -0.1em /\mathrm{c}}}
\def\GeVc{\ensuremath{\mathrm{Ge\kern -0.1em V}\kern -0.1em /\mathrm{c}}}
\def\MeVc{\ensuremath{\mathrm{Me\kern -0.1em V} /\mathrm{c}}}

\def\GeVcc{\ensuremath{\mathrm{Ge\kern -0.1em V}\kern -0.1em /\mathrm{c}^2}}
\def\MeVcc{\ensuremath{\mathrm{Me\kern -0.1em V}\kern -0.1em /\mathrm{c}^2}}
\def\eVcc{\ensuremath{\mathrm{e\kern -0.1em V}\kern -0.1em /\mathrm{c}^2}}

\def\cm{\ensuremath{\mathrm{cm}}}

\begin{document}

\title{Measurement of the $\WW$ cross section in $\sqrt{s} = 7~\TeV\ pp$ collisions with ATLAS}
\author{ATLAS Collaboration}
\preprint{CERN-PH-EP-2011-054; submitted to PRL}

\begin{abstract} 

This Letter presents a measurement of the $\WW$ production cross section 
in $\sqrt{s} = 7~ \TeV\ pp$ collisions by the ATLAS experiment, using 34~$\pbi$ 
of integrated luminosity produced by the Large Hadron Collider at CERN. 
Selecting events with two isolated leptons, each either an electron or a 
muon, 8 candidate events are observed
with an expected background of $1.7\pm0.6$ events. The measured 
cross section is $41^{+20}_{-16}(\mathrm{stat.})\pm 5(\mathrm{syst.})\pm 1(\mathrm{lumi.})$ pb, which 
is consistent with the standard model prediction of $44 \pm 3$~pb calculated 
at next-to-leading order in QCD. 

\end{abstract}


\maketitle

The $\WW$ process plays an important role in electroweak physics. The production rate and kinematic distributions of $\WW$ are sensitive to the triple gauge couplings of the $W$ boson~\cite{ref:lep_ww,ref:tev_ww} and $\WW$ production is an important background to standard model Higgs boson searches.
For these reasons, the measurement of the $\WW$ production cross section in 7 \TeV $pp$ collisions is a milestone in the Large Hadron Collider (LHC) physics program. $\WW$ production has been previously measured in both $e^+e^-$ collisions~\cite{ref:lep_ww} and $\ppbar$ collisions~\cite{ref:tev_ww}, and was also recently measured in $pp$ collisions~\cite{ref:cms_ww}. 
In the standard model, the largest production mechanisms of $\WW$ proceed via $s$-channel and $t$-channel quark annihilation, followed by the gluon fusion process, which is next-to-next-to-leading order but is enhanced by the large gluon-gluon parton luminosity at the LHC. 

Candidate $\WW$ events are reconstructed in the leptonic $\ell\ell\nu\nu$ decay channel where each $\ell$ is either an electron or a muon; included in this selection is $\WW$ production in which either or both $W$ bosons decay to $\tau\nu\rightarrow\ell\nu\nu\nu$.
This channel provides a significantly better signal to background ratio than the semileptonic or hadronic channels.
Events consistent with $pp \rightarrow \WW + X \rightarrow \ell\ell\nu\nu + X$ are selected by requiring two reconstructed oppositely-charged leptons and a large transverse momentum imbalance due to the neutrinos, which escape detection. 
There are four main backgrounds, all of comparable size:
 (i)  $W$+jets production with a jet misidentified as a lepton; 
 (ii) Drell-Yan production, which includes $Z/\gamma^*\rightarrow \ell\ell$ where the observed momentum imbalance is due to mismeasurements and $Z/\gamma^*\rightarrow \tau\tau \rightarrow \ell\ell+4\nu$; 
 (iii) top production ($\ttbar$ and $Wt$), which also produces two $W$ bosons, but is not considered signal and is suppressed by vetoing candidates with jets; 
 (iv) other diboson processes, which include $WZ$ production decaying to $\ell\ell\ell\nu$ where one charged lepton is lost, $ZZ$ with one $Z$ decaying to charged leptons and one $Z$ decaying to neutrinos, and $W\gamma$ with the photon misidentified as an electron.

The ATLAS detector~\cite{detectorpaper} has a cylindrical geometry~\cite{coordinatesystem} and consists of an inner tracking detector surrounded by a 2~T superconducting solenoid, electromagnetic and hadronic calorimeters, and a muon spectrometer with a toroidal magnetic field. The inner detector provides precision tracking for charged particles for $|\eta|<2.5$. It consists of silicon pixel and strip detectors surrounded by a straw tube tracker that also provides transition radiation measurements for electron identification. The calorimeter system covers the pseudorapidity range $|\eta|< 4.9$. For $|\eta|<2.5$, the electromagnetic calorimeter is finely segmented and plays an important role in electron identification.
The muon spectrometer has separate trigger and high-precision tracking chambers covering $|\eta|<2.7$. 
The transverse energy $\Et$ is defined to be $E\sin\theta$, where $E$ is the energy associated with a calorimeter cell or energy cluster. Similarly, $\pt$ is the momentum component transverse to the beam line.

A three-level trigger system selects events to record for offline analysis. During the data-taking period, the selections for at least one electron or muon were made progressively stricter, culminating in an $\ET>$ 15 GeV single electron or $\pT> 13$  GeV single muon requirement. The results presented here use a data sample corresponding to 34~$\pbi$ collected during 2010, where
the subsystems described were operational.

The signal acceptance is determined from a detailed Monte Carlo simulation. The $\qqbar\rightarrow\WW$ signal is simulated up to next-to-leading order in QCD with \mcatnlo~\cite{MCatNLO} and the gluon fusion process is simulated with \ggtwoww~\cite{Binoth:2006mf}; the CTEQ6.6~\cite{CTEQ66} and  CTEQ6M~\cite{CTEQ6m} parton distribution functions (PDFs), respectively, are used. \herwig~\cite{herwig} is used to model $W$ leptonic decays, parton showers, and hadronization, and \Jimmy~\cite{Jimmy} is used to simulate the underlying event. 
The detector response simulation~\cite{simulation} is based on the \geant program~\cite{Geant4}.
For the table and figures in this Letter, the standard model expectation for the $\WW$ signal is normalized to $44 \pm 3$~pb, which is the sum of the quark annihilation (97\%) and gluon fusion (3\%) processes, as calculated by \mcatnlo and \ggtwoww.

The luminosity in a single bunch-crossing was sufficient to produce multiple collisions, observed as multiple vertices, in the same recorded event. The vertex with the largest sum $\pt^2$ of the associated tracks is selected as the primary vertex. Additional inclusive $pp$ collisions are also simulated to reproduce the vertex multiplicity observed in data.

Electrons are reconstructed from a combination of a track found in the inner detector and an electromagnetic calorimeter energy cluster with $|\eta| < 1.37$ or $1.52 < |\eta| < 2.47$ to avoid the transition region between the barrel and the end-cap electromagnetic calorimeters. Candidate electrons must satisfy the ``tight'' selection~\cite{atlas-winclusive}, which requires the following measured quantities to be consistent with those from a promptly produced electron: shower shape, ratio of energy deposited in the hadronic to electromagnetic calorimeters, inner-detector track quality, track-to-shower matching, ratio of calorimeter energy measurement to track momentum, and transition radiation in the straw tube tracker. 
The electron is required to be isolated such that the sum of $\Et$ for calorimeter energy in a cone of size $\Delta R \equiv \sqrt{{\Delta \eta}^2 + {\Delta \phi}^2 }$ of 0.3 around the electron is less than 6 GeV, excluding energy associated with the electron cluster.
The overall electron reconstruction and identification efficiency is measured from data using $W\rightarrow e\nu$ and $Z\rightarrow ee$ candidates. It varies from 78\% for the central region ($|\eta|<0.8$) to 64\% in the forward region ($2.0< |\eta| < 2.47$) with a statistical uncertainty of less than 0.4\% and a systematic uncertainty of 5\%~\cite{atlas-winclusive} averaged over rapidity. The systematic uncertainty is due to background uncertainties in the $W$ and $Z$ samples and the consistency of efficiencies derived from the two samples.

Muon candidates are formed by associating muon spectrometer (MS) tracks with inner detector (ID) tracks after accounting for energy loss in the calorimeter. 
A common transverse momentum is determined using a statistical combination of the two tracks and is required to have $|\eta|<2.4$. 
To reject muons from charged $\pi$ or $K$ decays and charged particles 
from the beam-induced backgrounds, the MS muon $\pt$
must exceed 10~GeV and be consistent with the ID measurement, $|\pt^{\rm ID}-\pt^{\rm MS}|/\pt^{\rm ID} < 0.5 $. To suppress muons originating from hadronic jets, the sum of the $\pt$ of other tracks with $\pt>1$ GeV in a cone of $\Delta R$ = 0.2 around the muon candidate divided by the muon $\pt$ is required to be less than 0.1.
The muon reconstruction and isolation efficiencies are measured in data using $Z \rightarrow \mu\mu$ candidates to obtain a combined efficiency of $92\pm1(\mathrm{stat})\pm1(\mathrm{syst})$\%, where the systematic uncertainty is dominated by variations between data-taking periods due to additional collisions in the events~\cite{atlas-winclusive}. 

Jets used to discriminate top from $\WW$ production are reconstructed from calorimeter clusters using the anti-$k_{\mathrm{T}}$ algorithm~\cite{antikt-jet} with a resolution parameter of $R=0.4$. Jets within a $\Delta R<0.3$ of an electron are not used because the electrons are in general also reconstructed as jets.  The jet energies are calibrated using $\et$- and $\eta$-dependent correction factors~\cite{ref:Jet_Xsec} based on simulation and validated by test beam and collision data. 

In order to suppress the Drell-Yan background, a momentum imbalance of the visible collision 
products in the plane transverse to the beam axis is required. For this purpose,
missing transverse energy is defined as \mbox{$\metvec=-\sum\vecEt$}, where $\vecEt$ indicate the 2-dimensional transverse vectors for the reconstructed clusters of energy in the calorimeter in the range $|\eta| < 4.5$ and muon momenta. Since the $\metvec$ variable is sensitive to the mismeasurement of an individual lepton or jet, the relative missing transverse energy is defined as
\begin{equation*}
\label{eqn:MetRelDef}
\metrel = \left\{ 
\begin{array} {ll}
\met \times \sin\left(\Delta\phi\right) & \mbox{if} \; \Delta\phi < \pi/2 \\
\met                                    & \mbox{if} \; \Delta\phi \geq \pi/2,
\end{array}
\right.
\end{equation*}
\noindent where $\Delta\phi$ is the difference in the azimuthal 
angle $\phi$ between \metvec and the nearest lepton or jet.
This definition allows events to be removed when \metvec points along a lepton direction,
which occurs when the lepton momentum is measured lower than the true value or, for events with the two leptons moving in approximately opposite directions, higher than the true value.
This generally reduces the contribution from mismeasured leptons giving a higher signal to background ratio than a direct requirement on $|\metvec|$. Two important cases are high-mass muonic Drell-Yan events, where the momentum resolution can be comparable to $|\metvec|$ in $\WW$ events, and  $Z \rightarrow \tau\tau$, where the real \metvec from leptonic $\tau$ decays is parallel to the momenta of the leptons.

\begin{figure*}[t]
 \begin{center}
 \includegraphics[clip=true,trim=20 5 20 20,width=0.32\textwidth]{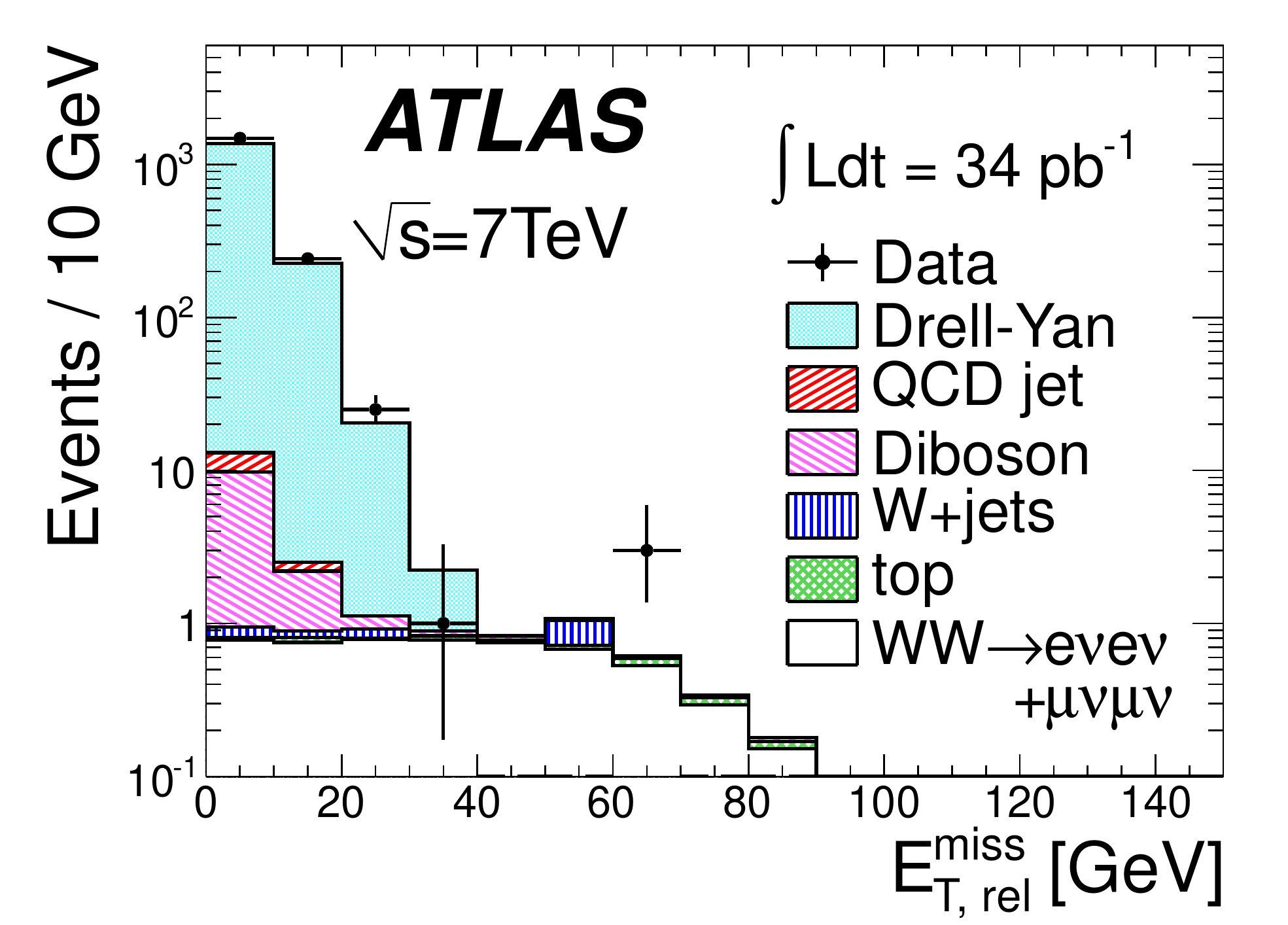}
 \includegraphics[clip=true,trim=20 5 20 20,width=0.32\textwidth]{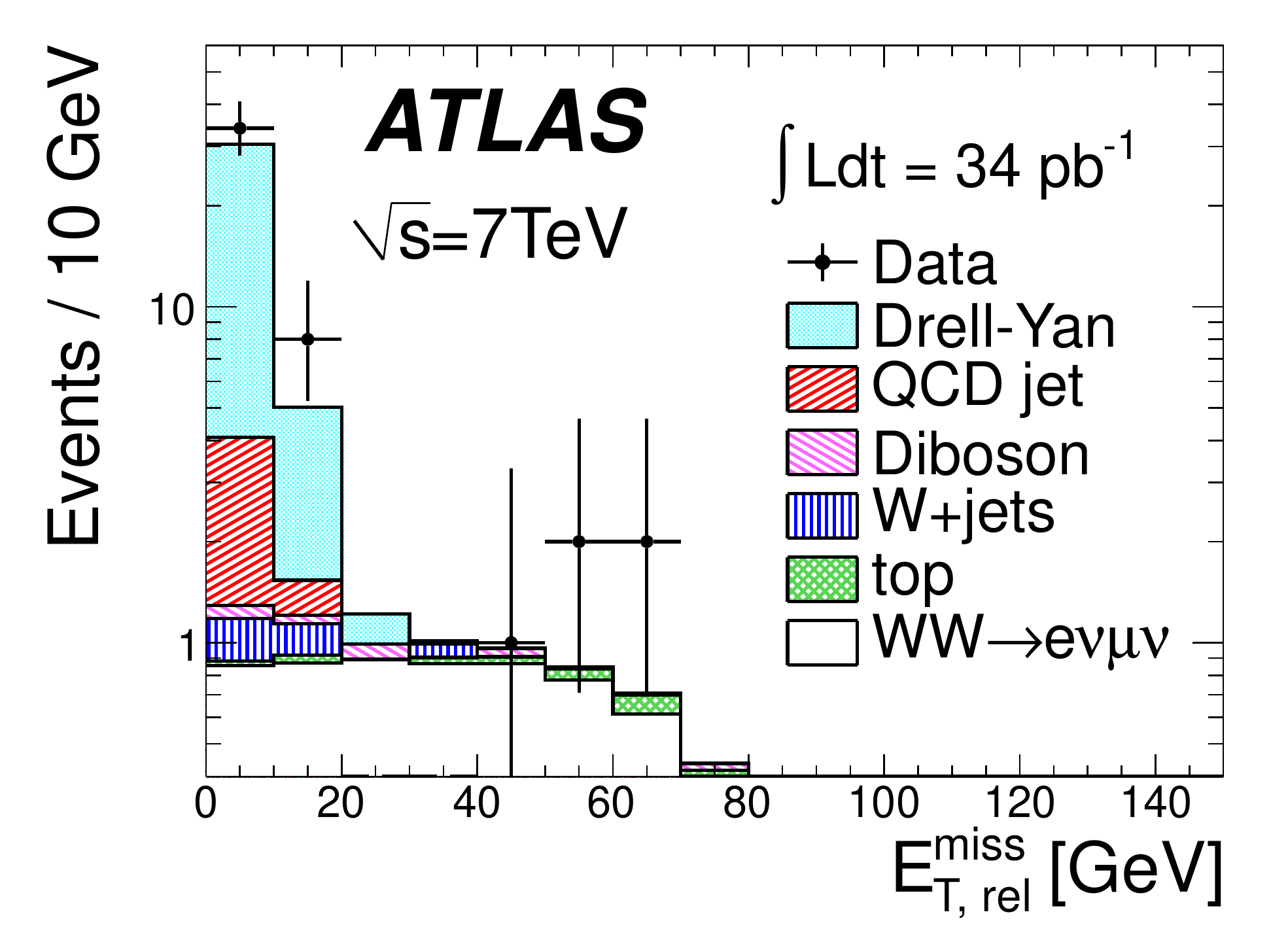}
 \includegraphics[clip=true,trim=20 5 20 20,width=0.32\textwidth]{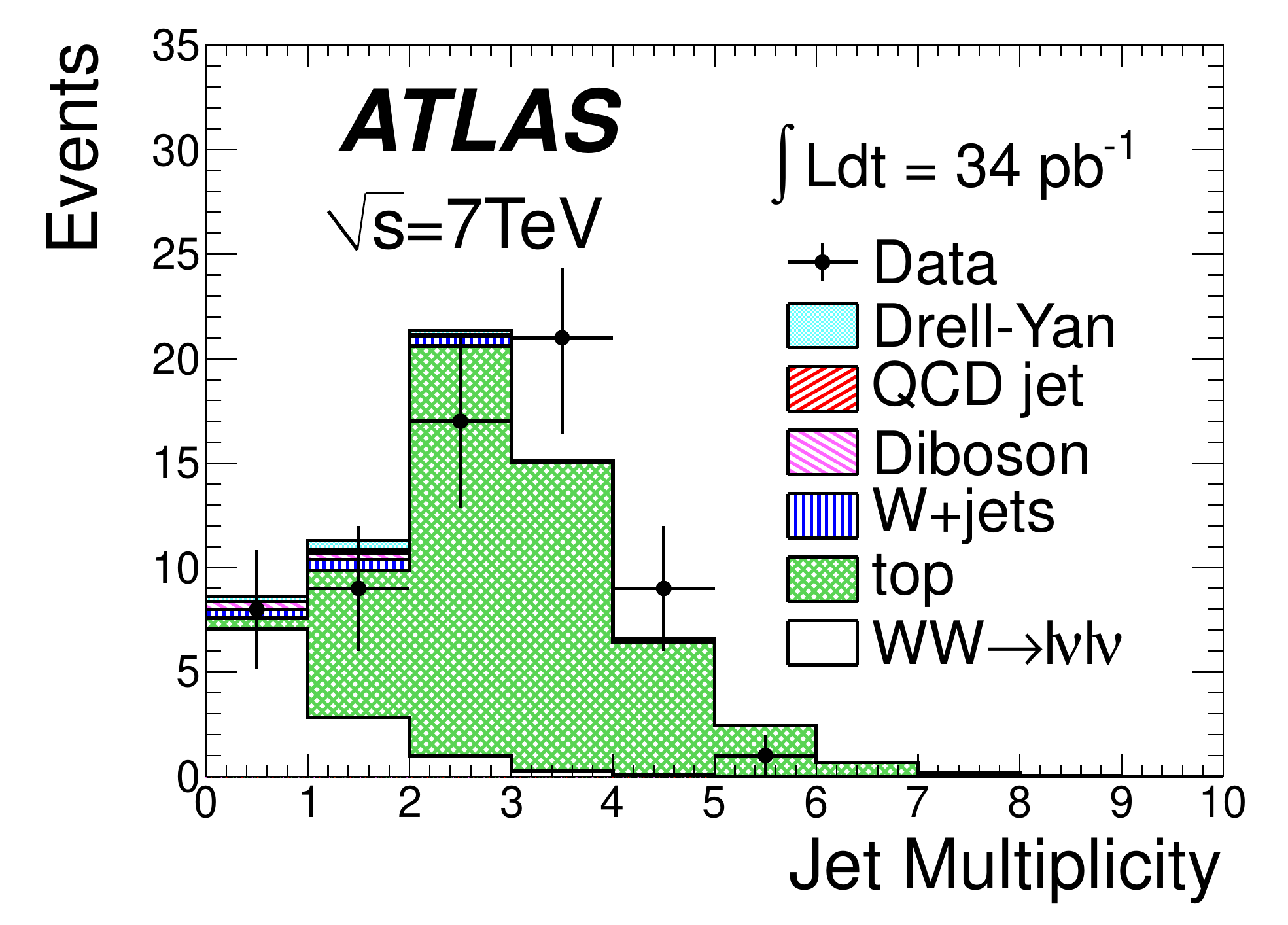}
 \caption{\metrel distributions for the selected $ee$ and $\mu\mu$ (left) and $e\mu$ (center) events and the multiplicity distribution for jets with $\pt>20$ GeV and $|\eta|<3.0$ for all three dilepton channels combined (right). The distributions show events with all selection criteria applied except for \metrel in the \metrel distribution and the jet veto in the jet multiplicity distribution. Simulation is used for the QCD jet and $W$+jets background contributions in these plots as opposed to the data-driven method used for $W$+jets in the signal region described in the text. The QCD jet contribution is negligible in the signal region.
 }
 \label{figs}
 \end{center}
 \end{figure*}

Candidates are selected with two opposite-sign charged leptons with $\pt>20$ GeV. The leptons are required to be consistent with coming from a primary vertex with at least three associated tracks, which makes the cosmic ray background negligible. For the $ee$ and $\mu\mu$ final states, the resulting sample is dominantly $Z\rightarrow ee$ and $Z\rightarrow \mu\mu$ events, while $e\mu$ candidates are a mix of $Z\rightarrow\tau\tau$,  $\ttbar$,  and other backgrounds. In order to suppress $Z/\gamma^*\rightarrow ee$ and $Z/\gamma^*\rightarrow \mu\mu$, $ee$ and $\mu\mu$ events with an invariant mass near the $Z$ mass, $|\mll-m_{Z}|<10$ GeV,  or $\mll<$ 15 GeV are removed and the remainder are required to have $\metrel>40$ GeV. For $e\mu$ events, a less stringent requirement $\metrel>20$ GeV is made.  In order to suppress the $\ttbar$ contribution, candidates containing jets with $\pt>20$~GeV and $|\eta|<3$ are removed. Figure~\ref{figs} shows the \metrel distributions separately for same-flavor ($ee$ and $\mu\mu$) and $e\mu$ events with all selections applied except for the \metrel requirement; also shown is the number of  jets with $\pt>20$ GeV after final selection except for the jet veto requirement. In all three distributions, the data in the background regions are in agreement with the standard model expectation and the signal is clearly visible.

\unitlength=\textwidth

The acceptance, which converts the observed yield in the kinematically restricted signal region to the inclusive $\WW$ cross section, is derived from simulation and is corrected with scale factors based on measurements in independent data samples.
The scale factors correct for the difference in trigger, lepton reconstruction and identification, and jet-veto efficiencies between data and simulation. The efficiency to pass the trigger criteria is close to unity and has small statistical and systematic uncertainties. 
For the lepton reconstruction and identification, the scale factors differ from unity by at most a few percent, indicating the accuracy of the simulation, and have systematic uncertainties derived from the efficiency measurements described above. A small smearing is added to the muon $p_T$ in simulation, so that it replicates the $Z\rightarrow\mu\mu$ invariant mass distribution in data.
The acceptance uncertainty due to the PDF uncertainties is 1.2\%.

 There are two major sources of systematic uncertainty in the jet-veto efficiency. The first is the modeling of jet production in association with $\WW$ due to initial state radiation, radiation from the internal line in the $t$-channel diagram, and additional parton collisions in the same $pp$ collision.
The second is the jet-energy scale, which is the correspondence between the true particle jet $\pt$ and the reconstructed jet $\pt$. To minimize the systematic uncertainty due to these two effects, control samples of $Z\rightarrow \ell\ell$ are used. These are sufficiently large and pure that the jet-veto efficiency can be directly measured and compared to simulation using the same QCD modeling as the $\WW$ signal. The ratio of the observed to the simulated zero-jet fraction in the $Z\rightarrow \ell\ell$ sample to simulation is used to define a jet-veto scale factor of $0.97 \pm 0.06.$ The uncertainty is due to differences between the jet-veto efficiency in $Z$ and $\WW$ events which is assessed including effects from the choice of renormalization and fragmentation scales \cite{Campbell:2009kg}. 

The overall selection acceptances for signal events are 
$4.1 \pm 0.1$\% for $e\nu e\nu$, $8.6 \pm 0.1$\% for $\mu\nu\mu\nu$, and $11.5 \pm 0.6$\% for $e\nu\mu\nu$. 
The relative acceptance in event selection  
are lepton acceptance and identification (18\%, 41\%, 27\%) 
and the $m_{ll}$ (85\%, 84\%, 100\%), $\metrel$ (41\%, 43\%, 69\%), 
and jet-veto (64\%, 59\%, 61\%) requirements, 
where the three percentages indicate the $ee$, $\mu\mu$, and $e\mu$ channels, respectively, and each factor is relative to the previous requirement.
The contributions from $\WW$ production where one or both $W$ bosons decays to a $\tau$ which subsequently
decays to an $e$ or $\mu$
are less than 10\% of the final selected \WW signal events in all three channels.

\begin{table*}
  \caption{Summary of observed events and expected standard model
           signal and background contributions
           in the three dilepton channels and their combination.
           The first uncertainty is statistical, the second systematic.}
  \begin{tabular}{l@{\hspace{0.8cm}}c@{\hspace{0.8cm}}c@{\hspace{0.8cm}}c@{\hspace{0.8cm}}c} \hline\hline
     Final State  & $e^+e^- \metrel$ & $\mu^+\mu^- \metrel$ & $e^\pm\mu^\mp \metrel$ & Combined \\

    \hline
    Observed Events  & 1 & 2 & 5 & 8 \\

    \hline

     Expected \WW             & $0.79 \pm 0.02 \pm 0.09$   & $1.61 \pm 0.04 \pm 0.14$   & $4.45 \pm 0.06 \pm 0.44$
                              & $6.85 \pm 0.07 \pm 0.66$ \\

     \hline
     \multicolumn{5}{l}{Backgrounds} \\

     ~~Drell-Yan              & $0.00 \pm 0.10 \pm 0.07$  & $0.01 \pm 0.10 \pm 0.07$   & $0.22 \pm 0.06 \pm 0.15$   
                              & $0.23 \pm 0.15 \pm 0.17$ \\

     ~~$WZ$, $ZZ$, $W\gamma$  & $0.05 \pm 0.01 \pm 0.01$  & $0.10 \pm 0.01 \pm 0.01$   & $0.23 \pm 0.05 \pm 0.02$   
                              & $0.38 \pm 0.04 \pm 0.04$ \\

     ~~$W+$jets               & $0.08 \pm 0.05 \pm 0.03$  & $0.00 \pm 0.29 \pm 0.10$   & $0.46 \pm 0.12 \pm 0.17$   
                              & $0.54 \pm 0.32 \pm 0.21$ \\

     ~~Top                    & $0.04 \pm 0.02 \pm 0.02$  & $0.14 \pm 0.06 \pm 0.07$   & $0.35 \pm 0.10 \pm 0.19$   
                              & $0.53 \pm 0.12 \pm 0.28$ \\

\hline
     Total Background         & $0.17 \pm 0.11 \pm 0.08$  & $0.25 \pm 0.31 \pm 0.15$   & $1.26 \pm 0.17 \pm 0.31$   
                              & $1.68 \pm 0.37 \pm 0.42$ \\

     \hline \hline
  \end{tabular}
  \label{ta:selected_data_MC}
\end{table*}

With the exception of $W$+jets, the backgrounds are derived from
simulations, corrected with the same scale factors as applied
to the modeling of the signal acceptance. The backgrounds are
scaled to the data sample based on the integrated luminosity and predicted cross sections. 
The top   
and $WZ$  
processes are simulated with \mcatnlo, 
the $ZZ$ 
process is simulated with \herwig,
the $W\gamma$ 
is simulated with {\sc madgraph}+\pythia~\cite{pythia,madgraph}, and 
the Drell-Yan process 
is simulated with \alpgen~\cite{alpgen} and \pythia~\cite{pythia}.
The QCD jet contribution, which is not significant after the $\metrel$ cut, is modeled with \pythia in Figure~\ref{figs}. 

Like the signal acceptance, the background estimates have uncertainties due to the trigger, lepton reconstruction and identification, and jet-veto efficiencies, in addition to the uncertainties on the integrated luminosity and theoretical cross sections. The Drell-Yan and top background estimates have additional uncertainties described below. 
Most of the Drell-Yan  events are removed by the dilepton invariant mass and \metrel requirements, but because of the large cross section some remain as background.  The uncertainty on this background due to the simulation of \metrel is assessed using a control sample of $Z/\gamma^* \rightarrow ee$ and $Z/\gamma^* \rightarrow \mu\mu$ events in the $Z$ mass peak region, $|\mll-m_{Z}|<10$ GeV, passing a relaxed requirement of $\metrel>$ 30 GeV. Despite the \metrel requirement, this sample is still dominated by $Z\rightarrow \ell\ell$ events in which the observed momentum imbalance is due to a combination of detector resolution, limited detector coverage, and neutrinos from heavy flavor decays. A 64\% systematic uncertainty is assigned based on the difference between the observed yield in data and the Monte Carlo prediction, which are statistically consistent. 

The top background arises from $\ttbar$ and $Wt$ production where the two $W$ bosons decay leptonically.
Simulation based on \mcatnlo is used to estimate the number of events passing the jet-veto requirement.
Similar to the signal acceptance, there are two important systematic uncertainties on the jet-veto efficiency: the jet-energy scale and the amount of initial and final state radiation (ISR/FSR). The jet-energy
calibration uncertainty~\cite{ref:Jet_Xsec} corresponds to a 40\% change in the top background.
The uncertainty due to the ISR/FSR modeling is estimated using the 
{\sc AcerMC}~\cite{Kersevan:2004yg} generator interfaced to \pythia~\cite{pythia}, and varying the 
parameters controlling ISR and FSR in a range consistent with experimental data~\cite{Skands:2010ak}.
The resulting uncertainty of 32\% is a combination of the shift in the prediction and the statistical uncertainty on the simulation.

$W$ bosons produced in association with a jet that is misidentified as a lepton
contribute to the selected sample.
The rate at which hadronic jets are misidentified as leptons may not be accurately 
described in the simulation, because these events are due to rare fragmentation processes or interactions with the detector. This background is therefore determined from data
using control samples dominated by $W$+jets events and subtracting all other components using simulation. The {\sc alpgen}+{\sc herwig}+{\sc jimmy} simulation of the $W$+jets background used in Figure 1 gives comparable results to this method. The $W$+jets data samples are constructed by requiring one electron or muon passing the full selection criteria and a lepton-like jet, which is a reconstructed electron or muon that is selected as likely to be due to a jet. For electrons, the lepton-like jets are 
electromagnetic clusters matched to tracks in the inner detector that fail the full
electron selection. For muons, lepton-like jets are muon candidates that fail at least one of the requirements on isolation, distance from the primary vertex, or ID and MS consistency. These events are otherwise required to pass the full event selection, treating the lepton-like jet as if it were a fully identified lepton.

The $W+$jets background is then estimated by scaling this control sample by a measured \pt-dependent factor $f$. The factor $f$ is the ratio of the probability for a jet to satisfy 
the full lepton identification criteria to the probability to satisfy the lepton-like jet criteria. 
The factor $f$ is measured in a QCD jet data sample and corrected for the small contribution of true leptons to the sample using simulation. 
The systematic uncertainty on $f$ is 36\% for both electrons and muons, and 
is determined from variations of $f$ in different run periods 
and in data samples containing jets of different energies, which covers 
differences in the quark/gluon composition between the jets in the QCD jet and $W +$ jets data samples.

\begin{figure}
 \begin{center}
 \includegraphics[clip=true,trim=2 0 5 15,width=0.155\textwidth]{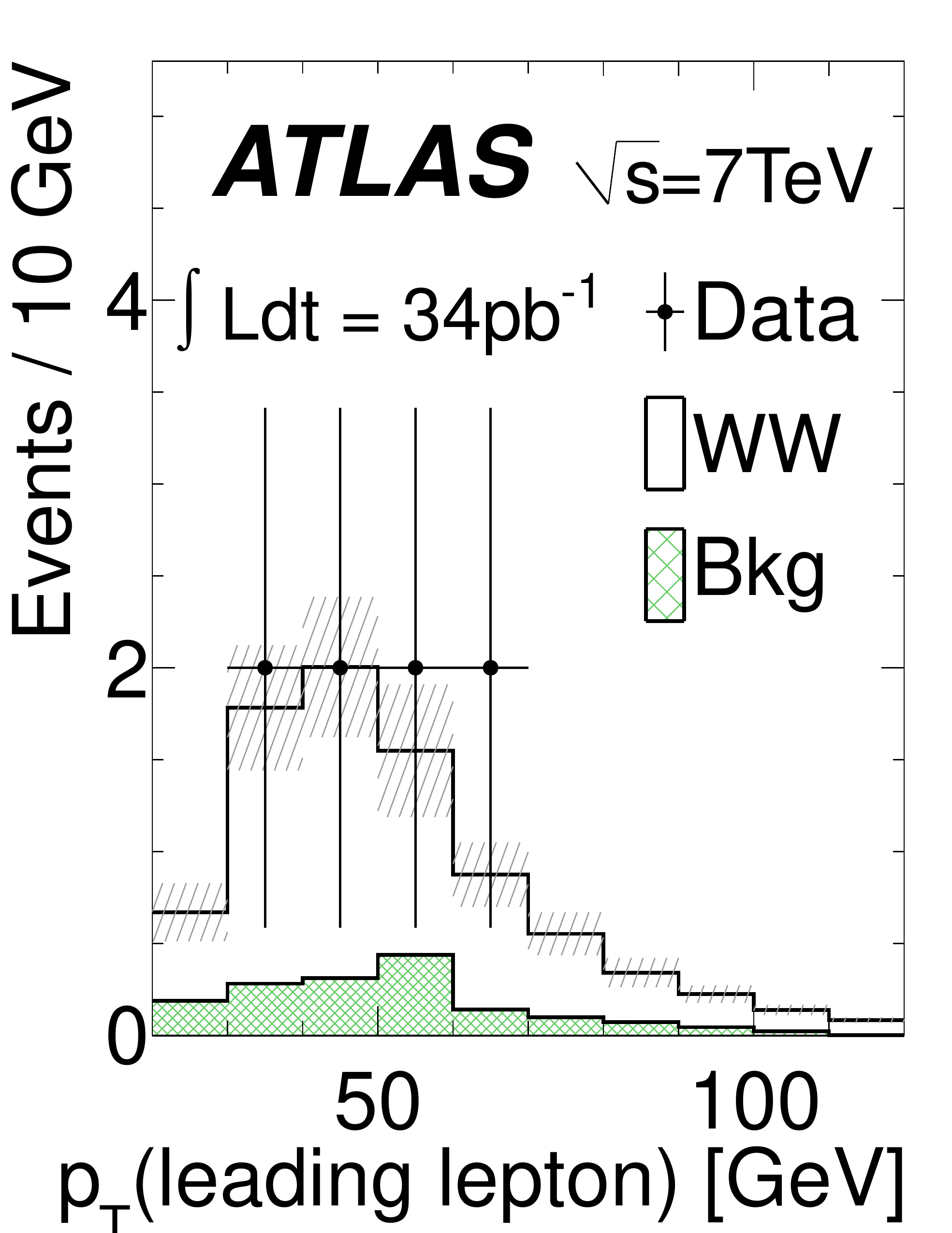}
 \includegraphics[clip=true,trim=2 0 5 15,width=0.155\textwidth]{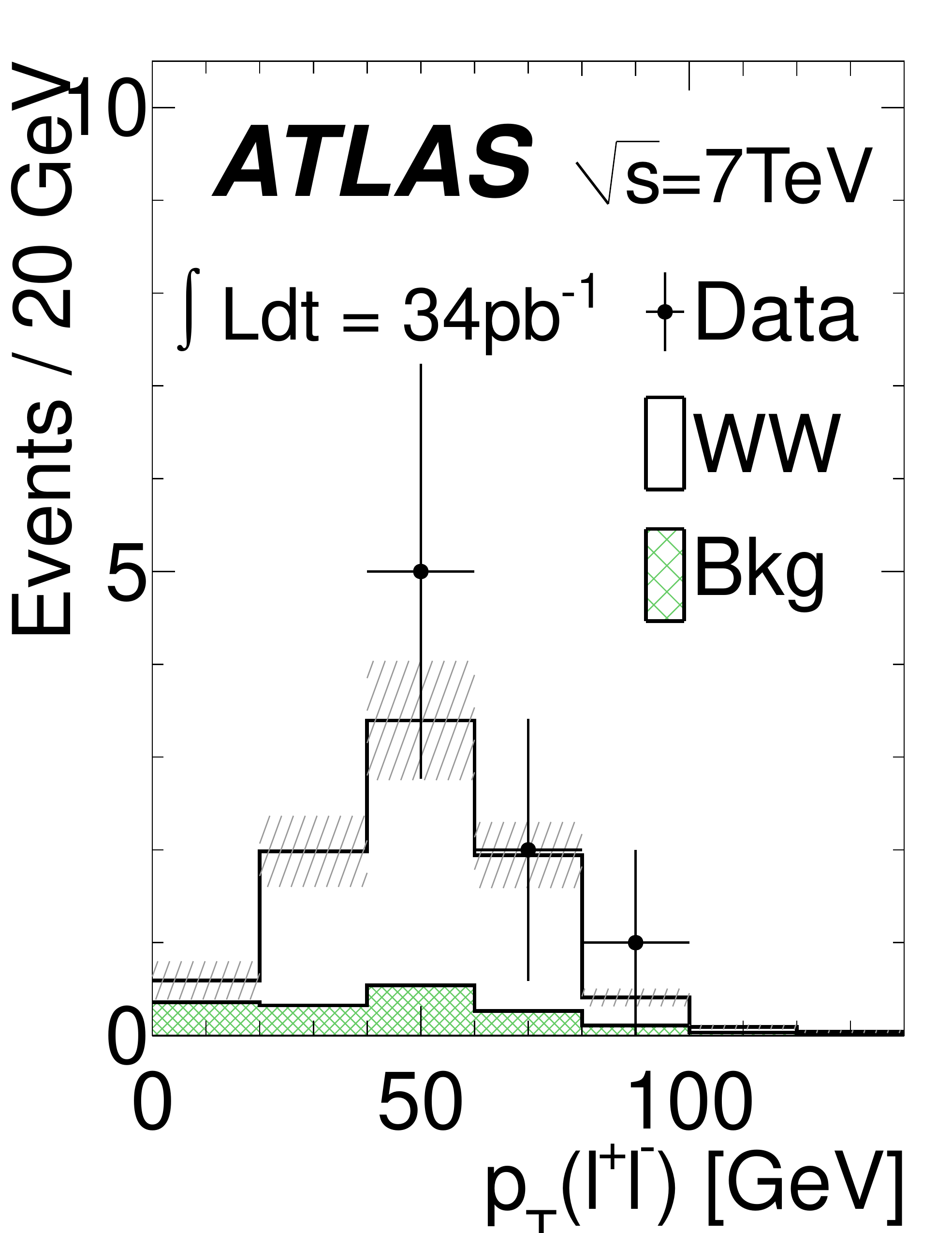}
 \includegraphics[clip=true,trim=2 0 5 15,width=0.155\textwidth]{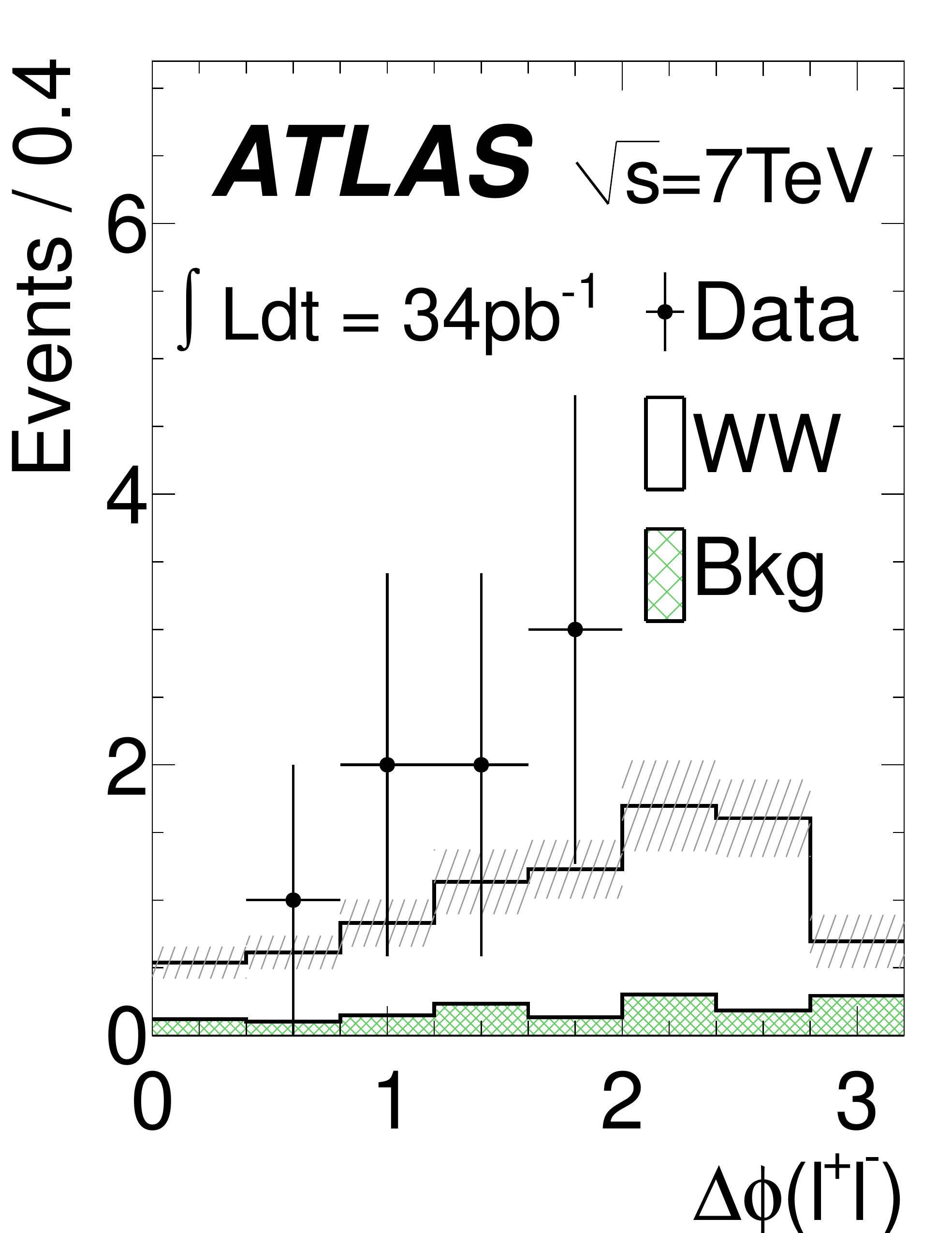}
 \caption{Distributions of the leading lepton $\pt$ (left), transverse momentum of the dilepton system (center), and azimuthal angle between the leptons (right) for the sum of the selected $ee$, $\mu\mu$ and $e\mu$ samples compared to the expectation. The gray band indicates the combined statistical and systematic uncertainty on the sum of the signal and background expectations.  }
 \label{fig:kinematics}
 \end{center}
 \end{figure}

The resulting signal and background expectations are shown in Table \ref{ta:selected_data_MC}. Eight events are observed in the data with a total expected background of $1.7\pm0.6$ events. As shown in Figure \ref{fig:kinematics}, the kinematic properties of the observed events are qualitatively consistent with the standard model expectation.  To estimate the statistical significance of the signal,  Poisson-distributed pseudo-experiments are generated, varying the expected background according to its uncertainty. The probability to observe 8 or more events in the absence of a signal is $1.2\times10^{-3}$, which corresponds to a significance of $3.0$ standard deviations. The \WW production cross section is determined using a maximum likelihood fitting method to combine the three dilepton channels. 
A cross section of 
$
\sigma_{\WW} \ = \ 41^{+20}_{-16}(\mathrm{stat.}) \pm 5(\mathrm{syst.}) \pm 1(\mathrm{lumi.})~\text{pb}
$
is measured.
The luminosity uncertainty for this measurement is 3.4\%~\cite{lumi}.
The total systematic uncertainty (11.5\%) includes the signal acceptance and efficiency ($\Delta A/A = 7.4\%$) 
and background estimation ($\Delta N_b/N_b = 33\%$) uncertainties. 
The dominant systematics uncertainties are due to the
jet-veto (7.5\%), 
and the lepton selection and identification (4.3\%) 

\unitlength=\textwidth

The measured $\WW$ production cross section is in good agreement with the standard model prediction of $44 \pm 3$~pb calculated at next-to-leading order in QCD and the recent measurement by CMS~\cite{ref:cms_ww}. With the significantly larger integrated luminosities expected to be provided by the LHC, this signal will form the basis of a research program that will include searches for the standard model Higgs boson, anomalous triple gauge couplings, and other processes beyond the standard model.

\section{Acknowledgements}

We thank CERN for the very successful operation of the LHC, as well as the
support staff from our institutions without whom ATLAS could not be
operated efficiently.

We acknowledge the support of ANPCyT, Argentina; YerPhI, Armenia; ARC,
Australia; BMWF, Austria; ANAS, Azerbaijan; SSTC, Belarus; CNPq and FAPESP,
Brazil; NSERC, NRC and CFI, Canada; CERN; CONICYT, Chile; CAS, MOST and
NSFC, China; COLCIENCIAS, Colombia; MSMT CR, MPO CR and VSC CR, Czech
Republic; DNRF, DNSRC and Lundbeck Foundation, Denmark; ARTEMIS, European
Union; IN2P3-CNRS, CEA-DSM/IRFU, France; GNAS, Georgia; BMBF, DFG, HGF, MPG
and AvH Foundation, Germany; GSRT, Greece; ISF, MINERVA, GIF, DIP and
Benoziyo Center, Israel; INFN, Italy; MEXT and JSPS, Japan; CNRST, Morocco;
FOM and NWO, Netherlands; RCN, Norway; MNiSW, Poland; GRICES and FCT,
Portugal; MERYS (MECTS), Romania; MES of Russia and ROSATOM, Russian
Federation; JINR; MSTD, Serbia; MSSR, Slovakia; ARRS and MVZT, Slovenia;
DST/NRF, South Africa; MICINN, Spain; SRC and Wallenberg Foundation,
Sweden; SER, SNSF and Cantons of Bern and Geneva, Switzerland; NSC, Taiwan;
TAEK, Turkey; STFC, the Royal Society and Leverhulme Trust, United Kingdom;
DOE and NSF, United States of America.

The crucial computing support from all WLCG partners is acknowledged
gratefully, in particular from CERN and the ATLAS Tier-1 facilities at
TRIUMF (Canada), NDGF (Denmark, Norway, Sweden), CC-IN2P3 (France),
KIT/GridKA (Germany), INFN-CNAF (Italy), NL-T1 (Netherlands), PIC (Spain),
ASGC (Taiwan), RAL (UK) and BNL (USA) and in the Tier-2 facilities
worldwide.

\begin{widetext}
\begin{flushleft}
{\Large The ATLAS Collaboration}

\bigskip

G.~Aad$^{\rm 48}$,
B.~Abbott$^{\rm 111}$,
J.~Abdallah$^{\rm 11}$,
A.A.~Abdelalim$^{\rm 49}$,
A.~Abdesselam$^{\rm 118}$,
O.~Abdinov$^{\rm 10}$,
B.~Abi$^{\rm 112}$,
M.~Abolins$^{\rm 88}$,
H.~Abramowicz$^{\rm 153}$,
H.~Abreu$^{\rm 115}$
% ...
\bigskip

% ...
$^{10}$ Institute of Physics, Azerbaijan Academy of Sciences, Baku, Azerbaijan\\
$^{11}$ Institut de F\'isica d'Altes Energies and Universitat Aut\`onoma  de Barcelona and ICREA, Barcelona, Spain\\
% ...
$^{48}$ Fakult\"{a}t f\"{u}r Mathematik und Physik, Albert-Ludwigs-Universit\"{a}t, Freiburg i.Br., Germany\\
$^{49}$ Section de Physique, Universit\'e de Gen\`eve, Geneva, Switzerland\\
% ...
$^{88}$ Department of Physics and Astronomy, Michigan State University, East Lansing MI, United States of America\\
% ...
$^{111}$ Homer L. Dodge Department of Physics and Astronomy, University of Oklahoma, Norman OK, United States of America\\
$^{112}$ Department of Physics, Oklahoma State University, Stillwater OK, United States of America\\
% ...
$^{115}$ LAL, Univ. Paris-Sud and CNRS/IN2P3, Orsay, France\\
% ...
$^{118}$ Department of Physics, Oxford University, Oxford, United Kingdom\\
% ...
$^{153}$ Raymond and Beverly Sackler School of Physics and Astronomy, Tel Aviv University, Tel Aviv, Israel\\
% ...
\end{flushleft}
\end{widetext}


\begin{thebibliography}{99}
\bibitem{ref:lep_ww}
       ALEPH Collaboration,    Phys.\ Lett.\  B {\bf 484}, 205 (2000);
       DELPHI Collaboration,   Phys.\ Lett.\  B {\bf 479}, 89 (2000);
  L3 Collaboration,
  Phys.\ Lett.\  B {\bf 496}, 19 (2000);
  OPAL Collaboration,
  Phys.\ Lett.\  B {\bf 493}, 249 (2000).

\bibitem{ref:tev_ww}
  CDF Collaboration,
  Phys.\ Rev.\ Lett.\  {\bf 104}, 201801 (2010);
  \D\O\ Collaboration,
  Phys.\ Rev.\ Lett.\  {\bf 103}, 191801 (2009).

\bibitem{ref:cms_ww}
  CMS~Collaboration,
  arXiv:hep-ex/1102.5429.

\bibitem{detectorpaper}
  ATLAS Collaboration, {JINST {\bf 3}, S08003} (2008).

\bibitem{coordinatesystem}
ATLAS uses a right-handed coordinate system with its origin at the nominal interaction point (IP) in the centre of the detector and the $z$-axis along the beam pipe. The $x$-axis points from the IP to the centre of the LHC ring, and the $y$ axis points upward. Cylindrical coordinates $(r,\phi)$ are used in the transverse plane, $\phi$ being the azimuthal angle around the beam pipe. The pseudorapidity is defined in terms of the polar angle $\theta$ as $\eta=-\ln\tan(\theta/2)$.

\bibitem{MCatNLO}
S.~Frixione and B.~R. Webber,  J. High Energy Phys. {\bf 0206},  029 (2002).

\bibitem{Binoth:2006mf} 
T.~Binoth, M.~Ciccolini, N.~Kauer, and M.~Kr\"amer,  J. High Energy Phys. {\bf 0612},  046 (2006).

\bibitem{CTEQ66}
P.~M. Nadolsky {\it et al.},  Phys. Rev. {\bf D78},  013004 (2008).

\bibitem{CTEQ6m}
  J.~Pumplin {\it et al.}, 
  J. High Energy Phys. {\bf 0207}, 012 (2002).

\bibitem{herwig}
G.~Corcella {\it et al.}, {J. High Energy Phys. {\bf 0101}, 010} (2001).

\bibitem{Jimmy}
J.~M. Butterworth {\it et al.}, {Z. Phys. {\bf C72},  637} (1996).

\bibitem{simulation}
{ATLAS} Collaboration, {Eur.~Phys.~J. {\bf C70}, 823 (2010)}.

\bibitem{Geant4} 
S.~Agostinelli {\it et al.}, {Nucl.Instrum.Meth.  {\bf A506}, 250 (2003)}.

\bibitem{atlas-winclusive}
{ATLAS} Collaboration,  {J. High Energy Phys. {\bf 1012},  060} (2010).

\bibitem{antikt-jet}
M.~Cacciari, G.~P. Salam, and G.~Soyez, {J. High Energy Phys. {\bf 0804},  063} (2008);
M.~Cacciari and G.~P. Salam, 
{Phys. Lett. {\bf B641} (2006)}.

\bibitem{ref:Jet_Xsec}
ATLAS Collaboration,
  Eur.\ Phys.\ J.\  C {\bf 71}, 1512 (2011).

\bibitem{Campbell:2009kg}
  J.~M.~Campbell  {\it et al.},
  Phys.\ Rev.\  {\bf D80}, 054023 (2009).

\bibitem{pythia}
  T.~Sjostrand, S.~Mrenna and P.~Skands,
  J. High Energy Phys. {\bf 0605}, 026 (2006).

\bibitem{madgraph}
J.~Alwall {\it et al.}, 
J. High Energy Phys.   {\bf 0709}  028, (2007)

\bibitem{alpgen}
M.~L.~Mangano {\it et al.},
J. High Energy Phys. {\bf 0307}, 001 (2003).

\bibitem{Kersevan:2004yg}
  B.~P.~Kersevan and E.~Richter-W\c{a}s, arXiv:hep-ph/0405247.

\bibitem{Skands:2010ak}
  P.~Z.~Skands,
  Phys.\ Rev.\  D {\bf 82}, 074018 (2010).

\bibitem{lumi}
ATLAS Collaboration,
ATLAS-CONF-2011-011, http://cdsweb.cern.ch/record/1334563.

\end{thebibliography}
\end{document}